\documentclass{ws-mpla}

\usepackage{amsmath}

\def\lsim{\raise0.3ex\hbox{$<$\kern-0.75em\raise-1.1ex\hbox{$\sim$}}}
\def\gsim{\raise0.3ex\hbox{$>$\kern-0.75em\raise-1.1ex\hbox{$\sim$}}}

\begin{document}

\baselineskip 6mm
\renewcommand{\thefootnote}{\fnsymbol{footnote}}

%------------ Hyun Seok's macro's, etc  -----------

\newcommand{\nc}{\newcommand}
\newcommand{\rnc}{\renewcommand}

\rnc{\baselinestretch}{1.24}    % 1.5 spacing btwn text lines
\setlength{\jot}{6pt}       % spacing btwn the rows of an eqnarray
\rnc{\arraystretch}{1.24}   % spacing btwn the rows of a non-eqn array

%%%%%%%%%%%%%%%%%%%%%% Equation Numbering %%%%%%%%%%%%%%%%%%%%%%%
\makeatletter
\rnc{\theequation}{\thesection.\arabic{equation}}
\@addtoreset{equation}{section}
\makeatother

%%%%%%%%%%%%%%%%%%%%%%%%%%%%%%%%%%%%%%%%%%%%%%%%%%%%%%%%%%%%%%%%%
%                                                               %
%                NEW COMMANDS AND MACROS                        %
%                                                               %
%%%%%%%%%%%%%%%%%%%%%%%%%%%%%%%%%%%%%%%%%%%%%%%%%%%%%%%%%%%%%%%%%

%%%%% Simplify some frequently used LaTeX commands %%%%%

\nc{\be}{\begin{equation}}

\nc{\ee}{\end{equation}}

\nc{\bea}{\begin{eqnarray}}

\nc{\eea}{\end{eqnarray}}

\nc{\xx}{\nonumber\\}

\nc{\ct}{\cite}

\nc{\la}{\label}

\nc{\eq}[1]{(\ref{#1})}

\nc{\what}[1]{\widehat{#1}}

\nc{\newcaption}[1]{\centerline{\parbox{6in}{\caption{#1}}}}

\nc{\fig}[3]{

\begin{figure}
\centerline{\epsfxsize=#1\epsfbox{#2.eps}}
\newcaption{#3. \label{#2}}
\end{figure}
}

%%% Caligraphic letters %%%%

\def\CA{{\cal A}}
\def\CC{{\cal C}}
\def\CD{{\cal D}}
\def\CE{{\cal E}}
\def\CF{{\cal F}}
\def\CG{{\cal G}}
\def\CH{{\cal H}}
\def\CK{{\cal K}}
\def\CL{{\cal L}}
\def\CM{{\cal M}}
\def\CN{{\cal N}}
\def\CO{{\cal O}}
\def\CP{{\cal P}}
\def\CR{{\cal R}}
\def\CS{{\cal S}}
\def\CU{{\cal U}}
\def\CV{{\cal V}}
\def\CW{{\cal W}}
\def\CY{{\cal Y}}
\def\CZ{{\cal Z}}

%%% Double line letters %%%

\def\IB{{\hbox{{\rm I}\kern-.2em\hbox{\rm B}}}}
\def\IC{\,\,{\hbox{{\rm I}\kern-.50em\hbox{\bf C}}}}
\def\ID{{\hbox{{\rm I}\kern-.2em\hbox{\rm D}}}}
\def\IF{{\hbox{{\rm I}\kern-.2em\hbox{\rm F}}}}
\def\IH{{\hbox{{\rm I}\kern-.2em\hbox{\rm H}}}}
\def\IN{{\hbox{{\rm I}\kern-.2em\hbox{\rm N}}}}
\def\IP{{\hbox{{\rm I}\kern-.2em\hbox{\rm P}}}}
\def\IR{{\hbox{{\rm I}\kern-.2em\hbox{\rm R}}}}
\def\IZ{{\hbox{{\rm Z}\kern-.4em\hbox{\rm Z}}}}

%%% Greek letters %%%

\def\a{\alpha}
\def\b{\beta}
\def\d{\delta}
\def\k{\kappa}
\def\l{\lambda}
\def\s{\sigma}
\def\t{\theta}
\def\w{\omega}
\def\G{\Gamma}

%%%%% Mathematical Symbols

\def\half{\frac{1}{2}}
\def\p{\partial}
\def\zbar{\bar{z}}

\markboth{Hyun Seok Yang}
{On the NCFT/Gravity Correspondence}
\catchline{}{}{}{}{}
%
%%%%%%%%%%%%%%%%%%%%%%%%%%%%%%%%%%%%%%%%%%%%%%%%%%%%%%%%%%%%%%%%%%%%

\title{On The Correspondence Between Noncommuative Field Theory And Gravity}

\author{\footnotesize Hyun Seok Yang}

\address{Institut f\"ur Physik, Humboldt Universit\"at zu Berlin,\\
Newtonstra\ss e 15, D-12489 Berlin, Germany \\
hsyang@physik.hu-berlin.de}

\maketitle
%\pub{Received (Day Month Year)}{Revised (Day Month Year)}

\begin{abstract}
In this brief review, I summarize the new development on the
correspondence between noncommuative (NC) field theory and gravity,
shortly referred to as the NCFT/Gravity correspondence. I elucidate
why a gauge theory in NC spacetime should be a theory of gravity. A
basic reason for the NCFT/Gravity correspondence is
that the $\Lambda$-symmetry (or $B$-field transformations) in NC spacetime
can be considered as a par with diffeomorphisms, which results from the
Darboux theorem. This fact leads to a striking picture about gravity:
Gravity can emerge from a gauge theory in NC spacetime. Gravity is then a collective
phenomenon emerging from gauge fields living in fuzzy spacetime.

\end{abstract}

\ccode{PACS Nos.: 11.10.Nx, 11.40.Gh, 04.50.+h}

\renewcommand{\thefootnote}{\arabic{footnote}}
\setcounter{footnote}{0}

\section{Introduction}

Gravity is a mysterious force: It is the weakest force in Nature,
but acts over great distances and is always attractive.
Therefore gravity is the final force to determine the evolution of our
Universe and is the origin causing strange phenomena, e.g., black
holes, in massive astronomical objects.
The special and the general relativity more mystify the gravity.
According to the theory of relativity, the space and the time are not independent concepts
but a single entity, the so-called spacetime and the matter and the energy can
be transformed into each other. These facts lead to a remarkable consequence
through the equivalence principle in general relativity that
gravitation arises out of the dynamics of spacetime being curved
by the presence of energy. So the gravity is quite different from
the electromagnetic force, the weak, and the strong nuclear forces,
that are all gauge theories regarding the spacetime as a background.

The origin of gravity as well as the quantum gravity, i.e., the
consistent dynamics of spacetime at a microscopic world, are still
out of reach. Fortunately, the string theory which has been recently
developed greatly provides us a consistent theory of quantum gravity
and reveals a remarkable and radical new picture about gravity. For
recent reviews, see, for example, Refs. [1-3]. String theory implies
that gravity may be not a fundamental force but a collective
phenomenon emerging from gauge theories such as a large $N$
Yang-Mills theory.

The general principle of quantum mechanics
implies that the structure of spacetime at a microscopic scale,
e.g., the Planck scale, is radically different from the continuous and
commutative spacetime familiar with our everyday life.
Rather the spacetime at the Planck scale is described by noncommutative (NC)
geometry, in other words, a fuzzy spacetime like the quantum mechanical world.
Quantum mechanics, which is the formulation of mechanics in NC phase space
($\hbar$-deformation), has revolutionarily changed classical physics, while NC
field theory, which is the formulation of field theory in NC spacetime
($\theta$-deformation), disappointingly has not revealed such a revolutionary
change up to now. For reviews on NC field theory, see Refs. [4,5].
Is the NC field theory a boring generalization or
seeding an unexplored revolution ?

I hope this review serves to reveal a revolutionary seed in NC field
theory, which has not been well appreciated, maybe, partially due to
a deep reluctance of physicists about NC field theory. (See the
Introduction in Ref. [4] for a comment about this socialogical
tendency.)
Throughout the recent works \ct{sty,ys,prl,prd},
we have arrived at a new understanding about the origin of gravity.
According to our results, the gravity is a collective phenomenon emerging
from electromagnetism in NC spacetime,
consistent with the picture also implied by string theory \ct{review2}.
It suggests that gravity is just the electromagnetism at the Planck
scale (as a typical scale where noncommutativity starts to work).
That is, the gravity and the electromagnetism at the Planck scale
are not independent forces but become an identical reality in the NC
spacetime. This is a completely new picture about the gravitation.
We expect that this picture will greatly promote our understanding
about the quantum nature of spacetime.

\section{Musing in Noncommutative Spacetime}

In order to provide a background for the correspondence between NC field theory and gravity,
shortly referred to as the NCFT/Gravity correspondence,
first I would like to present marked evidences
strongly indicating that NC field theories could be a theory
of gravity. In my opinion, this important physics has not been in a concrete
form although certain evidences are ubiquitous
from recent developments in string theory.
See Ref. [10] for a recent review along this line.
And then I briefly recapitulate the result in Ref. [9], showing that
NC electromagnetism should be a theory of gravity.
I will have minimal references in this review, but
see Ref. [9] for extensive literatures.

A NC spacetime is obtained by introducing a symplectic structure $B = \half
B_{\mu\nu} dy^\mu \wedge dy^\nu$ and then by quantizing the spacetime with
its Poisson structure $\theta^{\mu\nu} \equiv (B^{-1})^{\mu\nu}$,
treating it as a quantum phase space. That is, for $f,g \in C^\infty(M)$,
\be \la{poisson-bracket}
\{f, g\} = \theta^{\mu\nu} \left(\frac{\p f}{\p y^\mu} \frac{\p g}{\p y^\nu}
- \frac{\p f}{\p y^\nu} \frac{\p g}{\p y^\mu}\right) \Rightarrow
- i[\what{f},\what{g}].
\ee
According to the Weyl-Moyal map \ct{nc-review1,nc-review2}, the NC algebra of operators
is equivalent to the deformed algebra of functions defined
by the Moyal $\star$-product, i.e.,
\begin{equation}\label{star-product}
\what{f} \cdot \what{g} \cong (f \star g)(y) = \left.\exp\left(\frac{i}{2}
\theta^{\mu\nu} \partial_{\mu}^{y}\partial_{\nu}^{z}\right)f(y)g(z)\right|_{y=z}.
\end{equation}
Through the quantization rules \eq{poisson-bracket} and \eq{star-product},
one can define NC $\IR^4$ by the following commutation relation
\be \la{nc-spacetime}
[y^\mu, y^\nu]_\star = i \theta^{\mu\nu}.
\ee

{\bf I}. Although a field theory defined on the NC spacetime \eq{nc-spacetime}
preserves neither locality nor usual Lorentz invariance,
it was found \ct{twist-symm1,twist-symm2} that NC field theory is invariant
under the twisted Poincar\'e symmetry where the action of generators is now
defined by the twisted coproduct in the deformed Hopf algebras.
It is rather correct to say that the Lorentz symmetry is not broken
but realized as the deformed Hopf algebra and quantum group structures.
This symmetry plays a prominent role to construct
NC gravity \ct{nc-gravity1,nc-gravity2}. An important fact is that
translations in NC directions are basically gauge transformations, i.e.,
$e^{ik \cdot y} \star f(y) \star e^{-ik \cdot y} = f(y + k \cdot
\theta)$. This immediately implies that there are no local gauge-invariant
observables in NC gauge theory \cite{non-local1,non-local2}.
These properties are precisely those of gravity.
This was the motivation in Ref. [17] to explore the relation between NC field
theory and gravity.

{\bf II}. A NC field theory can be identified basically with a matrix model
or a large $N$ field theory. This claim is based on the following
fact. Let us consider a NC $\IR^2$ for simplicity:
\begin{equation} \label{nc-plane}
    [x,y] = i \theta.
\end{equation}
After scaling the coordinates $ x \to \sqrt{\theta} x, \; y \to \sqrt{\theta} y$,
the NC plane \eq{nc-plane} reduces to the Heisenberg algebra of
harmonic oscillator
\begin{equation} \label{oscillator}
    [a, a^\dagger] = 1.
\end{equation}
Therefore the representation space of NC $\IR^2$ is given by an
infinite-dimensional, separable Hilbert space $\CH = \{ |n \rangle, \; n=0,1,
\cdots \}$ and a scalar field $\widehat{\phi} \in \CA_\theta $ on the
NC plane \eq{nc-plane} can be expanded in terms of the complete operator basis
\begin{equation}\label{matrix-basis}
\CA_\theta = \{ |m \rangle \langle n|, \; n,m = 0,1, \cdots \},
\end{equation}
that is,
\begin{equation}\label{op-matrix}
    \widehat{\phi}(x,y) = \sum_{n,m} M_{mn} |m \rangle \langle n|.
\end{equation}
One can regard $M_{mn}$ in \eq{op-matrix} as components of an $N \times N$
matrix $M$ in the $N \to \infty$ limit.
More generally one may replace NC $\IR^2$ by a Riemann surface $\Sigma_g$
of genus $g$ which can be quantized via deformation quantization \ct{kontsevich}.
For a compact Riemann surface $\Sigma_g$ with finite area
$A(\Sigma_g)$, the matrix representation can be finite-dimensional.
In this case, $A(\Sigma_g) \sim \theta N$ but we simply take
the limit $ N \to \infty$.
We then arrive at the well-known relation:
\be \la{sun-sdiff}
\mathrm{Scalar \; field \; on \; NC} \; \IR^2 \;
(\mathrm{or} \; \Sigma_g) \Longleftrightarrow  N \times N \;
\mathrm{matrix} \; \mathrm{at} \; N \to \infty.
\ee
If $\widehat{\phi}$ is a real scalar field,
then $M$ should be a Hermitian matrix.
I argued in Ref. [9] that the above relation \eq{sun-sdiff} has far-reaching
applications to string theory.

The matrix representation \eq{op-matrix} clarifies why NC $U(1)$ gauge
theory is a large $N$ gauge theory. An important point is that the
NC gauge symmetry acts as unitary
transformations on $\CH$ for a field $\widehat{\phi} \in \CA_\theta $
in the adjoint representation of $U(1)$ gauge group
\be \la{ad-scalar}
\widehat{\phi} \to U  \widehat{\phi} \, U^\dagger.
\ee
This NC gauge symmetry $U_{\rm{cpt}}(\mathcal{H})$
is so large that $U_{\rm{cpt}}({\mathcal{H}}) \supset U(N)
\;(N \rightarrow \infty)$ \cite{nc-group1,nc-group2}, which is rather obvious in the
matrix basis \eq{matrix-basis}. Therefore the NC gauge theory
is essentially a large $N$ gauge theory. It becomes more explicit on
a NC torus through the Morita equivalence where NC
$U(1)$ gauge theory with rational $\theta = M/N$ is equivalent to an ordinary
$U(N)$ gauge theory \ct{morita1,morita2}. For this reason, it is not so surprising that NC
electromagnetism shares essential properties appearing in a large
$N$ gauge theory such as $SU(N \to \infty)$ Yang-Mills theory or matrix models.

{\bf III}. It is well-known \ct{thooft} that $1/N$ expansion
of any large $N$ gauge theory using the double line formalism reveals
a picture of a topological expansion in terms of surfaces of different
genus, which can be interpreted in terms of closed string variables
as the genus expansion of string amplitudes. It has been underlain
the idea that large $N$ gauge theories have a dual description in
terms of gravitational theories in higher dimensions. For example,
BFSS matrix model \ct{bfss}, IKKT matrix model \ct{ikkt}
and AdS/CFT duality \ct{ads-cft1,ads-cft2,ads-cft3}.
From the perspective \eq{sun-sdiff}, the $1/N$ expansion corresponds
to the NC deformation in terms of $\theta/A(\Sigma_g)$.

In particular, let us consider the NC electromagnetism in the
background independent formulation \ct{sw,seiberg}
since it is equivalent to the bosonic part of the IKKT matrix model
at large $N$ limit. The action for this case where $B_{\mu\nu} = (1/\theta)_{\mu\nu}$ is given by
\begin{eqnarray} \label{matrix-sw}
&& \frac{1}{4G_s}\int d^4 y (\widehat{F}-B)_{\mu\nu}
\star   (\widehat{F}-B)^{\mu\nu} \xx
&=& -\frac{\pi^2}{g_s \kappa^2} g_{\mu\lambda} g_{\nu\sigma}
{\mathbf{Tr}}_{\mathcal{H}} [x^\mu,x^\nu][x^\lambda, x^\sigma]
\end{eqnarray}
where we made a replacement $\frac{1}{(2\pi)^2} \int \frac{d^4y}{\mathrm{Pf}\theta}
\leftrightarrow  \mathbf{Tr}_{\mathcal{H}}$ using the Weyl-Moyal map \ct{nc-review1,nc-review2}.
The covariant and background-independent coordinates $x^\mu$ are
defined by \ct{seiberg,madore}
\be \la{cov-coord}
x^\mu(y) \equiv y^\mu + \theta^{\mu\nu} {\widehat A}_\nu(y)
\ee
and they are now regarded as operators on $\mathcal{H}$ which
is the representation space of the Heisenberg algebra \eq{nc-spacetime}.
The NC gauge symmetry $U_{\rm{cpt}}(\mathcal{H})$ in Eq.(\ref{matrix-sw})
then acts as unitary transformations on $\mathcal{H}$, i.e.,
\be \la{nc-symmetry}
x^\mu \rightarrow {x^{\prime}}^\mu = U x^\mu U^\dagger,
\ee
which is very large such that $U_{\rm{cpt}}({\mathcal{H}}) \supset
U(N) \;(N \rightarrow \infty)$ \cite{nc-group1,nc-group2}.
Note that the second expression of Eq.\eq{matrix-sw} in the matrix
basis \eq{matrix-basis} is equivalent to a large $N$ version of the IKKT
matrix model which describes the nonperturbative dynamics
of type IIB string theory \ct{ikkt}.

After pondering over the above observations {\bf I}-{\bf III}, one should raise a
question whether the electromagnetism in NC spacetime can be regarded
as a theory of gravity. Surprisingly, it turns out that the answer is YES !
I will briefly explain why NC electromagnetism should be a theory of
gravity, which has been referred in Refs. [8,9] to as the emergent gravity.

In order to understand the origin of the emergent gravity,
one has to identify the origin of diffeomorphism symmetry,
which is the underlying local symmetry of gravity.
It turned out \ct{prd} that the emergent gravity is deeply
related to symplectic geometry in sharp contrast to Riemannian geometry.
It can be shown \ct{future} that the emergent gravity is in general the
generalized complex geometry \ct{generalized-geometry1,generalized-geometry2}
and can be identified with the NC gravity \ct{nc-gravity1,nc-gravity2}
after full NC deformations.

The following intrinsic properties in the symplectic geometry \ct{book}
are crucial to understand the origin of the emergent gravity.

{\bf Symplectic manifold}: A symplectic structure on a smooth manifold $M$ is
a non-degenerate, closed 2-form $\omega \in \Lambda^2(M)$. The pair
$(M,\omega)$ is called a symplectic manifold. In classical mechanics,
the basic symplectic manifold is the phase space of $N$-particle
system with $\omega = \sum dq^i \wedge dp_i$.

{\bf Darboux theorem}: Locally, $(M,\omega) \cong (\IC^n, \sum dq^i \wedge
dp_i).$  That is, every $2n$-dimensional symplectic manifold can always
be made to look locally like the linear symplectic space $\IC^n$ with
its canonical symplectic form - Darboux coordinates.
This implies that, given two-forms $\omega$ and $\omega^\prime$
such that $[\omega]=[\omega^\prime] \in H^2(M)$,
then there exists a diffeomorphism $\phi: M \to M$ such that
$\phi^*(\omega^\prime)= \omega$.
In local coordinates, it is possible to find a coordinate transformation
$\phi$ whose pullback maps $\omega^\prime = \omega +dA$
to $\omega$, i.e., $\phi: y \mapsto x= x(y)$ so that
\begin{equation}\label{darboux}
    \frac{\partial x^\alpha}{\partial y^\mu} \frac{\partial x^\beta}{\partial
    y^\nu} \omega^\prime_{\alpha\beta}(x) = \omega_{\mu\nu}(y).
\end{equation}

The Darboux theorem leads to an important consequence on the low
energy dynamics of D-branes in the presence of a background
$B$-field. For a $Dp$-brane in arbitrary background fields,
the low energy dynamics is described by the Dirac-Born-Infeld (DBI)
action \ct{dbi1,dbi2} given by
\begin{equation} \label{dbi-general}
S = \frac{2\pi}{g_s (2\pi \kappa)^{\frac{p+1}{2}}}
\int d^{p+1} \sigma \sqrt{\det(g + \kappa (B + F))}
+ {\cal O}(\sqrt{\kappa} \partial F, \cdots),
\end{equation}
where $\kappa \equiv 2 \pi \alpha^\prime$, the size of a string, is a
unique expansion parameter to control derivative corrections.
The DBI action \eq{dbi-general} respects an important symmetry,
the so-called $\Lambda$-symmetry,
\be \la{lambda-symmetry}
B \to B - d\Lambda, \quad A \to A + \Lambda
\ee
for any one-form $\Lambda$. Thus the DBI action depends
on $B$ and $F$ only in the gauge invariant combination $\CF \equiv B
+ F$ as shown in \eq{dbi-general}. Note that ordinary $U(1)$ gauge
symmetry is a special case where the gauge parameters $\Lambda$ are
exact, namely, $\Lambda = d \lambda$, so that $B \to B, \; A \to A +
d\lambda$.

Suppose that the two-form $B$ is closed, i.e. $dB=0$, and
non-degenerate on the D-brane worldvolume $M$. The pair $(M, B)$
then defines a symplectic manifold.\footnote{\la{general-geometry} Note that
the `D-manifold' $M$ also carries a non-degenerate, symmetric,
bilinear form $g$ which is a Riemannian metric. The pair $(M,g)$
thus defines a Riemannian manifold.
If we consider a general pair $(M, g+ \kappa B)$, it
describes a generalized geometry \ct{generalized-geometry1,generalized-geometry2}
which continuously interpolates
between a symplectic geometry $(|\kappa Bg^{-1}| \gg 1)$ and
a Riemannian geometry $(|\kappa Bg^{-1}| \ll 1)$. The decoupling limit
considered in Ref. [29] corresponds to the former.
The symmetry \eq{lambda-symmetry} corresponds to $B$-field
transformations in Refs. [33,34].} But the
$\Lambda$-transformation \eq{lambda-symmetry} changes (locally) the
symplectic structure from $\omega = B$ to $\omega^\prime = B
-d\Lambda$. According to the Darboux theorem stated above,
there must be a coordinate transformation such
as Eq.\eq{darboux}. Thus the local change of symplectic structure
due to the $\Lambda$-symmetry can always be translated into
worldvolume diffeomorphisms as in Eq.\eq{darboux}.
That is, the Darboux theorem leads to an
interpretation of the $\Lambda$-symmetry as a diffeomorphism symmetry,
denoted as $G \equiv Diff(M)$, in the sense of Eq.\eq{darboux}.
Note that the number of gauge parameters in the $\Lambda$-symmetry is
exactly the same as $Diff(M)$. It turns out that the Darboux theorem
in symplectic geometry plays the same role as the equivalence principle
in general relativity.

The coordinate transformation in Eq.\eq{darboux} is not unique since
the symplectic structure remains intact if it is generated by a
vector field $X$ satisfying ${\cal L}_X B = 0$. Since we are
interested in a simply connected manifold $M$, i.e. $\pi_1(M)=0$,
the condition is equivalent to $\iota_X B + d \lambda = 0$, in other
words, $X \in Ham(M) \equiv$ the set of Hamiltonian vector fields.
Thus the symplectomorphism $H \equiv Ham(M)$
is equal to the $\Lambda$-symmetry where $\Lambda = d\lambda$ and
so $Ham(M)$ can be identified with the ordinary $U(1)$ gauge
symmetry \ct{cornalba,jurco-schupp}. For example, if a vector field
$X_\l$ is Hamiltonian satisfying $\iota_{X_\l} B + d \lambda = 0$,
the action of $X_\l$ on the covariant coordinates in \eq{cov-coord}
is given by
\bea \la{nc-gauge-tr}
\d x^\mu(y) &\equiv& X_\l (x^\mu) =
- \{\l, x^\mu \} \xx &=& \theta^{\mu\nu}(\p_\nu \l
+ \{ {\widehat A}_\nu, \l \}),
\eea
which is infinite dimensional as well as non-Abelian
and, after quantization \eq{poisson-bracket},
gives rise to NC gauge symmetry.

Using the $\Lambda$-symmetry, gauge fields can always be
shifted to $B$ by choosing the parameters as $\Lambda_\mu = - A_\mu$,
and the dynamics of gauge fields in the new symplectic form $B + dA$
is interpreted as a local fluctuation of symplectic structures. This
fluctuating symplectic structure can then be translated into a
fluctuating geometry through the coordinate transformation in
$G=Diff(M)$, the worldvolume diffeomorphism, modulo $H=Ham(M)$, the
$U(1)$ gauge transformation. We thus see that the `physical' change
of symplectic structures at a point in $M$ takes values in
$Diff_F(M) \equiv G/H = Diff(M)/Ham(M)$.

We need an explanation about the meaning of the `physical'.
The $\Lambda$-symmetry \eq{lambda-symmetry} is spontaneously broken to
the symplectomorphism $H = Ham(M)$ since the vacuum manifold
defined by the NC spacetime \eq{nc-spacetime} picks up
a particular symplectic structure, i.e.,
\be \la{spacetime-vacuum}
\langle B_{\mu\nu}(x) \rangle_{\rm{vac}} = (\theta^{-1})_{\mu\nu}.
\ee
This should be the case since we expect only the ordinary $U(1)$ gauge
symmetry in large distance (commutative) regimes, corresponding
to $|\kappa Bg^{-1}| \ll 1$ in the footnote \ref{general-geometry}
where $|\theta|^2 \equiv G_{\mu\l} G_{\nu\s}
\theta^{\mu\nu} \theta^{\l\s} = \kappa^2 |\kappa Bg^{-1}|^2 \ll \kappa^2 $
with the open string metric $G_{\mu\nu}$ defined by Eq.(3.21) in Ref. [29].
The fluctuation of gauge fields around the background \eq{spacetime-vacuum}
induces a deformation of the vacuum manifold, e.g. $\IR^4$ in the case of
constant $\theta$'s. According to the Goldstone's theorem,
massless particles, the so-called Goldstone bosons, should appear
which can be regarded as dynamical variables taking values
in the quotient space $G/H = Diff_F(M)$.

Since $G=Diff(M)$ is generated by the set of $\Lambda_\mu = - A_\mu$,
so the space of gauge field configurations on NC $\IR^4$ and $H=Ham(M)$
by the set of gauge transformations, $G/H$ can be identified with
the gauge orbit space of NC gauge fields, in other words, the `physical'
configuration space of NC gauge theory. Thus the moduli space of all possible
symplectic structures is equivalent to the `physical' configuration
space of NC electromagnetism. Note that the symmetry breaking \eq{spacetime-vacuum}
explains why gravity is physically observable in spite of
the {\it gauge symmetry} $Diff(M)$.

The Goldstone bosons for the spontaneous symmetry breaking $G \to H$ turn out
to be spin-2 gravitons, which might be elaborated by the following
argument. Using the coordinate transformation \eq{darboux} where $\w^\prime = B+F(x)$
and $\w = B$, one can get the following identity \cite{cornalba}
for the DBI action \eq{dbi-general}
\begin{equation} \label{proof-sw}
\int d^{p+1} x \sqrt{\det(g + \kappa (B + F(x)))}
= \int d^{p+1} y \sqrt{\det(\kappa B + h(y))},
\end{equation}
where fluctuations of gauge fields now appear as an induced metric on
the brane given by
\begin{equation} \label{induced-metric}
h_{\mu\nu}(y) =  \frac{\partial x^\alpha}{\partial y^\mu}
\frac{\partial x^\beta}{\partial y^\nu} g_{\alpha\beta}.
\end{equation}
The dynamics of gauge fields is then encoded into the fluctuations
of geometry through the embedding functions $x^\mu(y)$ defined by
\eq{cov-coord}. As usual, $y^\mu$ are vacuum expectation
values of $x^\mu$ specifying the background \eq{spacetime-vacuum} and
$\what{A}_\mu(y)$ are fluctuating (dynamical) coordinates (fields).

The above argument clarifies why the dynamics of NC gauge fields can be
interpreted as the fluctuations of geometry described by
the metric (\ref{induced-metric}). We may identify
$\partial x^\alpha/\partial y^\mu \equiv e^\a_\mu (y)$ with vielbeins on some
manifold $\CM$ by regarding $h_{\mu\nu}(y)= e^\a_\mu(y) e^\b_\nu(y) g_{\a\b}$
as a Riemannian metric on $\CM$.
The embedding functions $x^\mu(y)$ in \eq{cov-coord}, which are now dynamical
fields, subject to the equivalence relation, $x^\mu \sim x^\mu + \d x^\mu$,
defined by the gauge transformation \eq{nc-gauge-tr}, coordinatize the
quotient space $G/H = Diff_F(M)$.
In this context, the gravitational fields $e^\a_\mu (y)$ or $h_{\mu\nu}(y)$
correspond to the Goldstone bosons for the spontaneous symmetry breaking
\eq{spacetime-vacuum}. This is a rough picture showing how gravity
can emerge from NC electromagnetism. I refer Ref. [9] for more details.

\section{Deformation Quantization and Seiberg-Witten Map}

Deformation quantization is a deformation of the commutative
product in the algebra $\CA$ of classical
observables which consist of a Poisson manifold to a NC, associative product.
In a seminal paper [18], M. Kontsevich proved that every
finite-dimensional Poisson manifold $M$ admits a canonical deformation quantization.
The deformation quantization provides a noble approach to reify the Darboux
theorem beyond a semi-classical, i.e. $\CO(\theta)$, limit.
In the context of deformation quantization, the Darboux theorem appears as
an automorphism of $\CA[[\hbar]]$ considered as
an $\IR[[\hbar]]$--module (i.e. linear transformations $\CA \to \CA$
parameterized by $\hbar$). If $D(\hbar) = 1+ \sum_{n\geq{1}}{\hbar^{n}D_n}$
is such an automorphism where $D_n: \CA \to \CA$ are differential operators,
it acts on the set of star products as
\be \la{star-equiv}
\star \rightarrow \star^\prime, \quad
f(\hbar) \star^\prime g(\hbar) = D(\hbar)
\Bigl( D(\hbar)^{-1} (f(\hbar)) \star  D(\hbar)^{-1} (g(\hbar)) \Bigr)
\ee
for $f(\hbar), g(\hbar) \in \CA[[\hbar]]$.

It was shown in Ref. [18] ({\it Theorem 1.1} and {\it Theorem 2.3}) that
the set of gauge equivalence classes of star products on a smooth manifold $M$
can be naturally identified with the set of equivalence classes of Poisson
structures modulo the action of the diffeomorphism group of $M$, starting
at the identity diffeomorphism and, if we change coordinates,
we obtain a gauge equivalent star product.

Note that the gauge equivalence \eq{star-equiv} is defined up to the
following inner automorphism \ct{jurco-schupp,jusch-wess}
\be \la{inner-autom}
f(\hbar) \to \l(\hbar) \star f(\hbar) \star \l(\hbar)^{-1}
\ee
or its infinitesimal version is
\be \la{inner-auto-0}
\d f(\hbar) = i [ \l, f ]_\star.
\ee
The above similarity transformation \la{inner-auto} definitely does not
change star products. The inner automorphism \eq{inner-autom} is equivalent
to the NC gauge transformation \eq{nc-symmetry}, which is the quantum
deformation of Eq.\eq{nc-gauge-tr}.

In summary, the $\Lambda$-symmetry \eq{lambda-symmetry} (see
the footnote \ref{general-geometry}) is realized as
the gauge equivalence \eq{star-equiv} between star products
and the $U(1)$ gauge symmetry appears as the inner
automorphism \eq{inner-autom}, which is the NC $U(1)$ gauge symmetry.

If we make an arbitrary change of coordinates, $y^\mu \mapsto
x^\mu(y)$, in the Moyal $\star$-product \eq{star-product}
which is nothing but Kontsevich's star product with the
constant Poisson bi-vector $\theta^{\mu\nu}$,
we get a new star product defined by a Poisson bi-vector $\alpha(\hbar)$.
But the resulting star product has to be gauge equivalent
to the Moyal product \eq{star-product} and $\alpha(\hbar)$ should be
determined by the original Poisson bi-vector $\theta^{\mu\nu}$.
This was explicitly checked by Zotov in Ref. [41] and
he obtained the deformation quantization formula up to the third order.

The equivalence \eq{star-equiv} under the change of coordinates, $y^\mu \mapsto
x^\mu(y)$, immediately leads to the exact Seiberg-Witten (SW) map \ct{prd}
\be \la{esw-deformation}
[x^\mu,x^\nu]_{\star} = i(\theta - \theta \what{F}(y) \theta)^{\mu\nu}
= 2 D(\hbar)^{-1}(\alpha^{\mu\nu})
\ee
where the left hand side is the Moyal product \eq{star-product} and the new Poisson structure
is given by
\be \la{esw-inverse}
\a^{\mu\nu}(x) = \frac{i}{2}\left(\frac{1}{B+F}\right)^{\mu\nu}(x).
\ee
The explicit expressions of $D(\hbar)$ and $\a^{\mu\nu}$ are given in the
Refs. [9,41]. Thus the exact SW map can be found by either Eq.\eq{esw-deformation} or
Eq.\eq{esw-inverse} if it is determined how the map  $D(\hbar)$ or the Poisson structure
$\a^{\mu\nu}$ depends on the coordinate transformation,
as was done up to $\CO(\hbar^2)$ in Ref. [41].

Incidently, we would like to point out that the gauge equivalence
\eq{star-equiv} between star products is an equivalent statement with the SW
equivalence \ct{sw} between commutative and NC DBI actions, in general,
including derivative corrections too, as was explained in Ref. [9].
See also Refs. [39,40].

\section{Noncommutative Instantons and Twistor Space}

The NCFT/Gravity correspondence claimed in section 2 can be beautifully
confirmed, at least, for the self-dual sector of NC gauge theories \ct{sty,ys,prl,prd}.
It turns out that self-dual NC electromagnetism perfectly fits with the
twistor space describing curved self-dual spacetime \ct{penrose,atiyah}.

Let us consider electromagnetism in the NC spacetime defined by Eq.\eq{nc-spacetime}.
The action for the NC $U(1)$ gauge theory in flat Euclidean $\IR^4$
is given by
\begin{equation}\label{nced}
\widehat{S}_{\mathrm{NC}} = \frac{1}{4}\int\! d^4 y \,\widehat{F}_{\mu\nu}
\star \widehat{F}^{\mu\nu}.
\end{equation}
Contrary to ordinary electromagnetism, the NC $U(1)$ gauge theory admits
non-singular instanton
solutions satisfying the NC self-duality equation \ct{nek-sch},
\be \la{nc-self-dual}
{\widehat F}_{\mu\nu} (y) = \pm \half
\varepsilon_{\mu\nu\lambda\sigma}{\widehat F}_{\lambda\sigma} (y).
\ee

The NC gauge theory \eq{nced} has an
equivalent dual description through the SW map in terms of ordinary gauge
theory on commutative spacetime \ct{sw}.
For simplicity, we will be confined to semi-classical limit, say $\CO(\theta)$
in Eq.\eq{star-product}, which means slowly varying fields,
$\sqrt{\kappa} |\frac{\partial F}{F}| \ll 1$,
in the sense keeping field strengths (without restriction on their
size) but not their derivatives.
In this limit, the Moyal-Weyl commutator in Eq.\eq{esw-deformation}
can be replaced by the Poisson bracket \eq{poisson-bracket} and
$D(\hbar) \approx 1$.

Applying the map \eq{esw-deformation} with \eq{esw-inverse}, we get the exact
SW map in the semi-classical limit \ct{hsy,ban-yang}
\be \la{semi-esw}
\what{F}_{\mu\nu}(y) = \left(\frac{1}{1+F\theta} F
\right)_{\mu\nu}(x).
\ee
Using the map \eq{semi-esw} together with the measure transformation
\be \label{sw-measure}
d^{p+1} y = d^{p+1} x \sqrt{\det(1+ F \theta)}(x),
\ee
one can get the commutative nonlinear electrodynamics \ct{hsy,ban-yang}
equivalent to Eq.\eq{nced} in the semi-classical
approximation,
\begin{equation}\label{ced-sw}
S_{\mathrm{C}} = \frac{1}{4} \int d^4 x \sqrt{\det{{\rm g}}} \;
{\rm g}^{\mu \lambda} {\rm g}^{\sigma\nu} F_{\mu\nu}
F_{\lambda\sigma},
\end{equation}
where we introduced an effective metric ${\rm g}_{\mu\nu}$ induced by
dynamical gauge fields as follows
\begin{equation}\label{effective-metric}
    {\rm g}_{\mu\nu} = \delta_{\mu\nu} + (F\theta)_{\mu\nu},
    \qquad  ({\rm g}^{-1})^{\mu\nu} \equiv {\rm g}^{\mu\nu} =
    \Bigl(\frac{1}{1 + F\theta}\Bigr)^{\mu\nu}.
\end{equation}
It was shown in Ref. [6] that the self-duality equation for the action
$S_{\mathrm{C}}$ is given by
\be \la{c-self-dual}
{\bf F}_{\mu\nu} (x) = \pm \half
\varepsilon_{\mu\nu\lambda\sigma}{\bf F}_{\lambda\sigma} (x),
\ee
where
\be \la{bold-f}
{\bf F}_{\mu\nu}(x) = \Bigl({\rm g}^{-1} F \Bigr)_{\mu\nu}(x).
\ee
Note that Eq.\eq{c-self-dual} is nothing but the exact SW map \eq{semi-esw}
of the NC self-duality equation \eq{nc-self-dual}.

I will show that the self-duality equation \eq{nc-self-dual} or
equivalently \eq{c-self-dual} describes gravitational instantons obeying
the self-duality equations \ct{g-instanton}
\be \la{g-instanton}
R_{abcd} = \pm \half \varepsilon_{abef}
{R^{ef}}_{cd},
\ee
where $R_{abcd}$ is a curvature tensor.
Let us rewrite ${\rm g}_{\mu\nu}$ as
\be \la{new-metric}
{\rm g}_{\mu\nu}= \half(\delta_{\mu\nu} + \widetilde{{\rm g}}_{\mu\nu})
\ee
and consider the line element defined by the metric $\widetilde{{\rm g}}_{\mu\nu}$
 with $\theta^{\mu\nu} = \frac{\zeta}{2} \eta^3_{\mu\nu}$
\be \la{dia-metric}
ds^2 = \widetilde{{\rm g}}_{\mu\nu} dx^\mu dx^\nu \equiv
 \widetilde{\sigma}_\mu \otimes  \widetilde{\sigma}_\mu.
\ee
It is easy to check \ct{ys} that $\widetilde{\sigma}_1 \wedge  \widetilde{\sigma}_2 \wedge
\widetilde{\sigma}_3 \wedge  \widetilde{\sigma}_4 = d^4 x$, in other words,
$\sqrt{\det{\widetilde{{\rm g}}_{\mu\nu}}} = 1$.
We then define the triple of K\"ahler forms as follows,
\be \la{unit-2-form}
\widetilde{\omega}^a = \half \eta^a_{\mu\nu} \widetilde{\sigma}^\mu
\wedge \widetilde{\sigma}^\nu, \qquad a=1,2,3.
\ee
One can easily see that
\bea \la{3-kahlerform}
&& \omega \equiv \widetilde{\omega}^2 + i \widetilde{\omega}^1 = dz_1 \wedge dz_2,
\quad \bar{\omega} \equiv \widetilde{\omega}^2 - i \widetilde{\omega}^1
=  d\zbar_1 \wedge d\zbar_2, \xx
&& \Omega \equiv - \widetilde{\omega}^3 = \frac{i}{2}(dz_1 \wedge d\zbar_1
+ dz_2 \wedge d\zbar_2) + \zeta F.
\eea
It is obvious that $d\widetilde{\omega}^a = 0, \; \forall a$. This means that
the metric $\widetilde{{\rm g}}_{\mu\nu}$ is hyper-K\"ahler \ct{ys}, which is
an equivalent statement as Ricci-flat K\"ahler in four dimensions.
Therefore the metric $\widetilde{{\rm g}}_{\mu\nu}$ is a gravitational
instanton \ct{g-instanton}.
Eq.\eq{3-kahlerform} clearly shows how dynamical gauge fields living in NC
spacetime induce a deformation of background geometry
through gravitational instantons, thus realizing the emergent geometry.

The deformation of symplectic (or K\"ahler) structure on $\IR^4$
due to the fluctuation of gauge fields can be more clarified by the
following construction, which closely follows the
result on $N=2$ strings \ct{ooguri-vafa}.
Consider a deformation of the holomorphic (2,0)-form
$\omega = dz_1 \wedge dz_2$ as follows
\be \la{deformation}
\Psi (t) = \omega + i t \Omega + \frac{t^2}{4} \bar{\omega}
\ee
where the parameter $t$ takes values in ${\bf S}^2$.
One can easily see \ct{prl,prd} that $d\Psi(t) = 0$ due to the Bianchi identity $dF=0$
and
\be \la{psi-psi}
\Psi (t) \wedge \Psi (t) = 0
\ee
since Eq.\eq{psi-psi} is equivalent to the instanton equation \eq{c-self-dual}.
Since the two-form $\Psi(t)$ is closed and degenerate, the Darboux theorem
asserts that one can find a $t$-dependent
map $(z_1,z_2)  \to (Z_1(t;z_i, \zbar_i), Z_2(t;z_i, \zbar_i))$ such that
\be \la{symplectic-form}
\Psi (t) = dZ_1(t;z_i,\zbar_i) \wedge dZ_2(t;z_i, \zbar_i).
\ee

When $t$ is small, one can solve \eq{symplectic-form} by expanding
$Z_i(t;z,\zbar)$ in powers of $t$ as
\be \la{small-expansion}
Z_i(t;z,\zbar) = z_i + \sum_{n=1}^{\infty} \frac{t^n}{n}p_n^i(z,\zbar).
\ee
By substituting this into Eq.\eq{deformation}, one gets at ${\cal O}(t)$
\bea \la{exp-eq1}
&& \partial_{z_i} p_1^i = 0, \\
\la{exp-eq2}
&& \epsilon_{ik} \partial_{\zbar_j} p_1^k dz^i \wedge d\zbar^j= i \Omega.
\eea
Eq.\eq{exp-eq1} can be solved by setting $p_1^i = 1/2 \epsilon^{ij}
\partial_{z_j} K$ and then $\Omega = i/2  \partial_i
{\bar \partial_j} K dz^i \wedge d\zbar^j$.
The real-valued smooth function $K$ is the K\"ahler
potential of $U(1)$ instantons \ct{sty,ys}.
In terms of this K\"ahler two-form $\Omega$, Eq.\eq{psi-psi} results
in the complex Monge-Amp\`ere or the Pleba\'nski equation \ct{prl,prd},
\be \la{cma-pleb}
\Omega \wedge \Omega = \half \omega \wedge \bar{\omega},
\ee
that is, $\det(\partial_i {\bar \partial_j} K) =1$.

When $t$ is large, one can introduce another Darboux coordinates
${\widetilde Z}_i(t;z_i,\zbar_i)$ such that
\be \la{symplectic-form2}
\Psi (t) = t^2 d{\widetilde Z}_1(t;z_i,\zbar_i) \wedge
d{\widetilde Z}_2(t;z_i, \zbar_i)
\ee
with expansion
\be \la{small-expansion2}
{\widetilde Z}_i(t;z,\zbar) = \zbar_i + \sum_{n=1}^{\infty} \frac{t^{-n}}{n}
{\widetilde p}_n^i(z,\zbar).
\ee
One can get the solution \eq{deformation} with ${\widetilde p}_1^i = - 1/2
\epsilon^{ij} \partial_{\zbar_j} K$ and $\Omega =
i/2  \partial_i {\bar \partial_j} K dz^i \wedge d\zbar^j$.

The $t$-dependent Darboux coordinates $Z_i(t;z,\zbar)$ and
${\widetilde Z}_i(t;z,\zbar)$ correspond to holomorphic coordinates
on two local charts, where the 2-form $\Psi(t)$ becomes the holomorphic
(2,0)-form, of the dual projective twistor space $\CZ$
as a fiber bundle over ${\bf S}^2$
with a fiber ${\cal M}$, a hyper-K\"ahler manifold.
We know from Eq.\eq{3-kahlerform} that $\Omega = dx^1 \wedge dx^2 + dx^3
\wedge dx^4 + \zeta(B+F)$ is rank 4 while $\w$ and $\bar{\w}$ are both rank 2.
Thus the K\"ahler form $\Omega$ can always serve as a symplectic form on
both coordinate charts and two sets of coordinates
at $t=0$ and $t=\infty$ are related to each other by a canonical
transformation given by $f_i(t;Z(t))=t {\widetilde Z}_i(t)$ \ct{ooguri-vafa}.
This fact leads to a beautiful result \ct{beautiful} that
the canonical transformation between them is generated by the K\"ahler
potential appearing in the complex Monge-Amp\`ere equation or
the Pleba\'nski equation.
In this way, the complex geometry of the twistor space $\CZ$ encodes all the
information about the K\"ahler geometry of self-dual 4-manifolds.

I argued in Ref. [9] that our approach here provides an intriguing recipe
for an inhomogeneous background, for example, specified by
\be \la{curved-spacetime}
\langle B^\prime_{\mu\nu}(x) \rangle_{\rm{vac}}
= ({\theta^\prime}^{-1})_{\mu\nu}(x).
\ee
If we regard $B^\prime_{\mu\nu}(x)$ as coming from an inhomogeneous
gauge field condensation on a constant $B_{\mu\nu}$ background, say,
$B^\prime_{\mu\nu}(x) = (B+ F_{\rm{back}}(x))_{\mu\nu}$,
a NC gauge theory with nonconstant NC parameters
${\theta^\prime}^{\mu\nu}(x)$ can be considered as that defined
by the usual Moyal star product \eq{star-product}
but around a nonperturbative solution described by $F_{\rm{back}}(x)$.
For instance, if $F_{\rm{back}}(x)$ describes NC instantons,
the vacuum manifold \eq{curved-spacetime} is a Ricci-flat K\"ahler manifold
and the corresponding NC gauge theory is defined
in the NC instanton background \ct{lee-yang}.
I think that this picture can shed some light
on NC gravity \ct{nc-gravity1,nc-gravity2}.

\section*{Acknowledgments}
We would like to thank Harald Dorn,
George Jorjadze and Jan Plefka for helpful discussions.
This work was supported by the Alexander von Humboldt Foundation.

\end{document}